 \documentclass[prb,twocolumn,showpacs,amsmath,amssymb,
                letterpaper,citeautoscript]{revtex4}

\usepackage[dvips]{graphicx}% Include figure files
\usepackage{dcolumn}% Align table columns on decimal point
\usepackage{bm}% bold math

\begin{document}

%\preprint{APS/123-QED}

\title{  
Contact Atomic Structure and Electron Transport Through Molecules 
}

\author{ San-Huang Ke,$^{1,2}$ Harold U. Baranger,$^{2}$ and Weitao Yang$^{1}$}

\affiliation{
     $^{\rm 1}$Department of Chemistry, Duke University, Durham, NC 27708-0354 \\
     $^{\rm 2}$Department of Physics, Duke University, Durham, NC 27708-0305
}

\date{\today}% It is always \today, today,

\begin{abstract}

Using benzene sandwiched between two Au leads as a model system, we investigate from first principles the change in molecular conductance caused by different atomic structures around the metal-molecule contact. 
Our motivation is the variable situations that may arise in break junction experiments; our approach is a combined density functional theory and Green function technique.
We focus on effects caused by (1) the presence of an additional Au atom at the contact and (2) possible changes in the molecule-lead separation. 
The effects of contact atomic relaxation and two different lead orientations are fully considered.
We find that the presence of an additional Au atom at each of the two contacts will increase the equilibrium conductance  by up to two orders of magnitude regardless of either the lead orientation or different group-VI anchoring atoms.
This is due to a LUMO-like resonance peak near the Fermi energy. 
In the non-equilibrium properties, the resonance peak manifests itself in a large negative differential conductance.
We find that the dependence of the equilibrium conductance on the molecule-lead separation can be quite subtle: either very weak or very strong depending on the separation regime.

\end{abstract}

\maketitle
%===================================================================
\section{Introduction}

Understanding electron transport through nanoscale junctions or molecular devices connected to metallic electrodes may be the basis of future molecular electronics technology \cite{mol1,mol2,mol3,mol4,mol5}.  One of the critical issues in this regard is to construct contact structures which can provide both useful stability and high contact transparency.  

In many recent experiments \cite{mol3,bj1,bj2,bj3}, Au electrodes were used as leads for transport measurements because of the high conductivity, stability, and well-defined fabrication techniques involved. A common way to construct a lead-molecule-lead (LML) system is by using a mechanically controllable break junction (MCB). These can be made either through direct mechanical means \cite{bj1,bj2,bj3,h2} or through electromigration \cite{morpurgo1,park,morpurgo2,fuhrer}. In these break-junction experiments the detailed atomic structure of the molecule-lead contacts of a LML system is unknown.  In fact, because of the atomic scale roughness of the break surface, different atomic scale structures of the contact may occur in different experiments.  Up to now, it has been difficult to investigate and control experimentally the detailed contact atomic structure and find out its influence on the electron transport through a molecular device. Here theoretical modeling/simulation free from empirical parameters can play an important role in understanding, interpreting observed experimental behaviors, or doing pre-designs for good contact structures.

Because of the need for an atomic scale description, {\it ab initio} density functional theory (DFT) \cite{dft1,dft2} is a natural approach to molecular electronics. One route works explicitly with scattering states via the Lippmann-Schwinger equation \cite{lang1,lang2,lang3,lang4}. The more common method, however, is to combine DFT using a localized basis set for molecular electronic structure with the non-equilibrium Green function method (NEGF) \cite{negf1,negf2} for electron transport \cite{datta1,datta2,transiesta,mcdcal,pal1,trank1}. In this method some parts of the electrodes of a LML system can be included in the device region to form an ``extended molecule'', and therefore the specific contact atomic structure and relaxation can be fully considered in principle although in practice the atomic structure is usually predetermined to avoid the heavy computational effort \cite{xue1,xue2,xue3,trans1,trans2,trans3,mcdcal1,mcdcal2,mcdcal3,mcdcal4,mcdcal5,pal2,pal3}.  In the implementations of this approach, some researchers \cite{datta1,datta2,xue1,xue2,xue3,pal1,pal2,pal3} adopted quantum chemistry methods for the DFT calculation in which a cluster geometry is used for the device region, while others \cite{mcdcal,mcdcal1,mcdcal2,mcdcal3,mcdcal4,mcdcal5,transiesta,trans1,trans2,trans3,trank1}
used a periodic geometry (as in solid state physics) for the device region.  The advantages of the latter are that the electronic structure of the device region and the two leads can be easily treated on the same footing and the infinite LML system is nearly perfect in geometry without any artificially introduced surface effect.

Previously we developed a self-consistent approach within the DFT+NEGF method for calculating electron transport through molecular devices \cite{trank1}.  Our approach is simple while strict: the non-equilibrium condition under a bias is fully included in the NEGF rather than the DFT part. Therefore, it is straightforward to combine with any electronic structure method that uses a localized basis set. More importantly, in this way the problem of solving for the Hartree potential under a bias field with unphysical potential jumps at the two boundaries of the DFT supercell is avoided.  In our method large parts of the two metallic leads of a system are included in the device region so that the molecule-lead interactions (including electron transfer and atomic relaxation) are fully included, and the electronic structures of the molecule and the two leads are treated {\it exactly} on the same footing.

%=================================== Fig. 1 ========================================
\begin{figure*}[htb]
\includegraphics[angle=  0,width= 6.0cm]{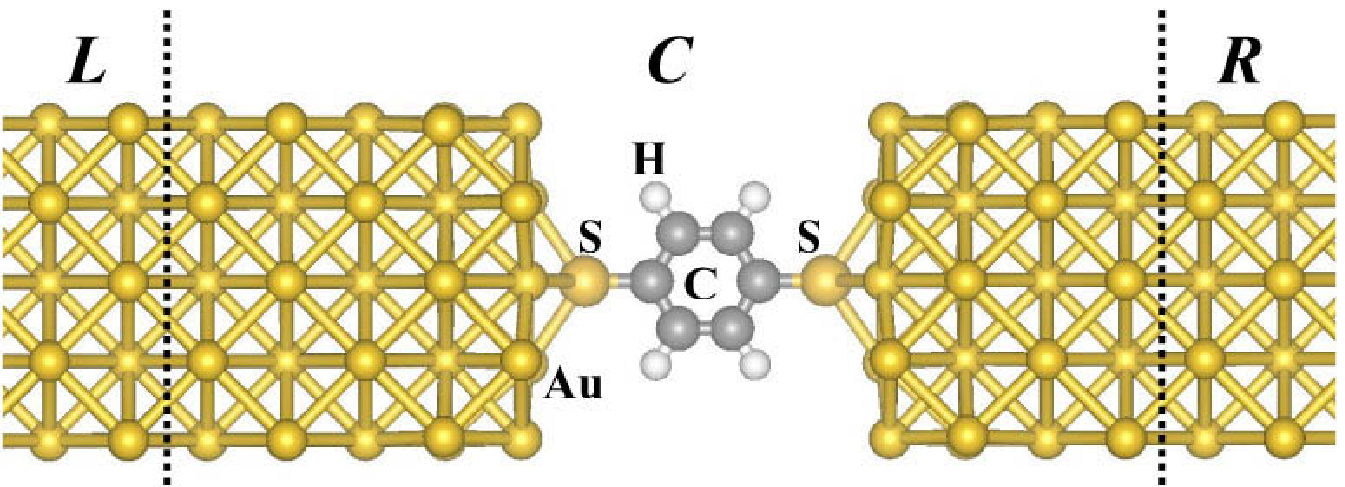} {\large \bf (a)}
\includegraphics[angle=  0,width= 7.0cm]{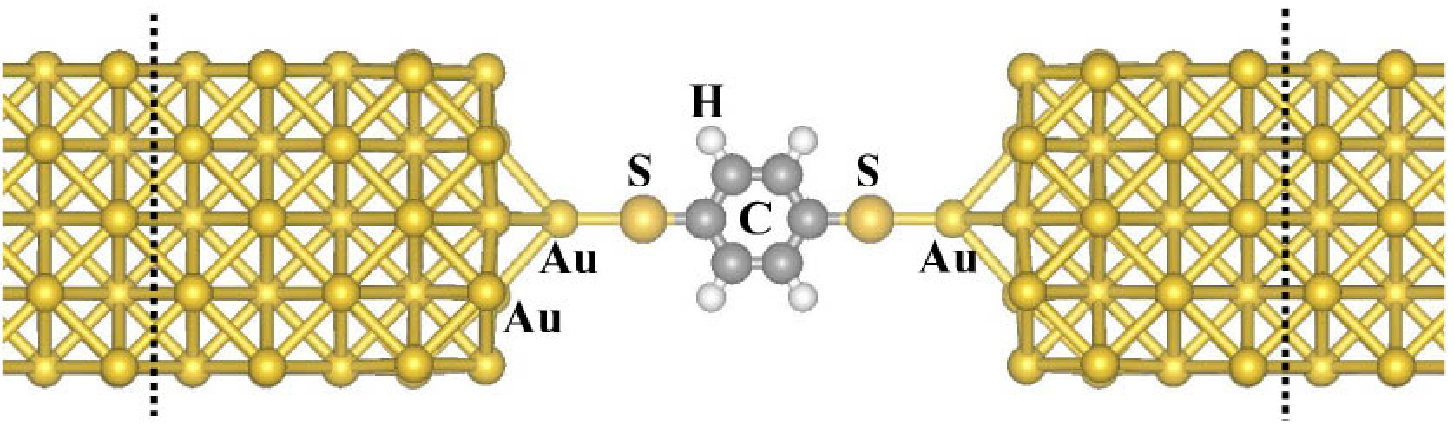} {\large \bf (b)} \\ 
\includegraphics[angle=  0,width= 6.5cm]{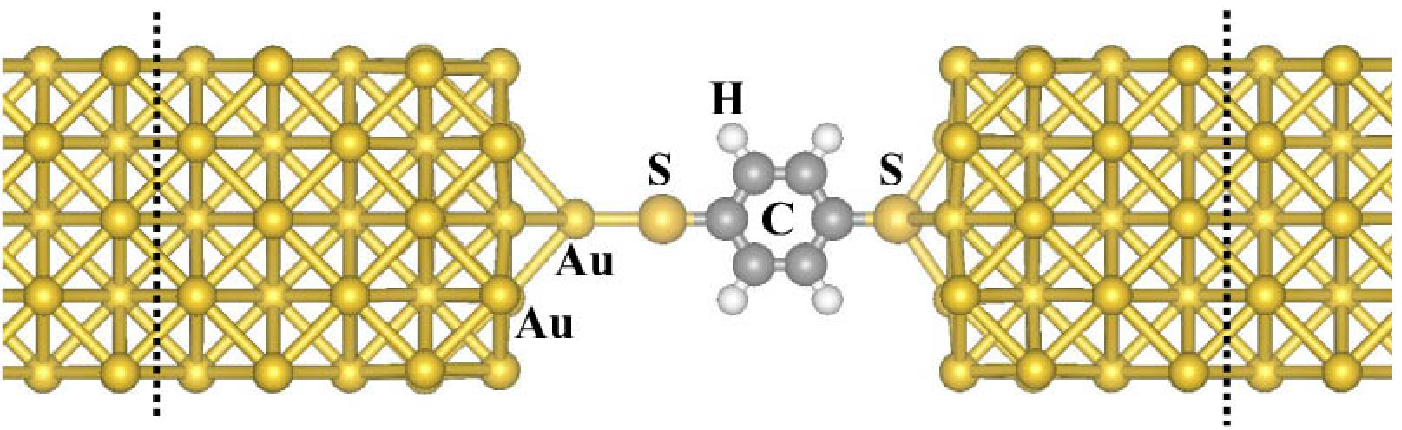} {\large \bf (c)}
\includegraphics[angle=  0,width= 6.8cm]{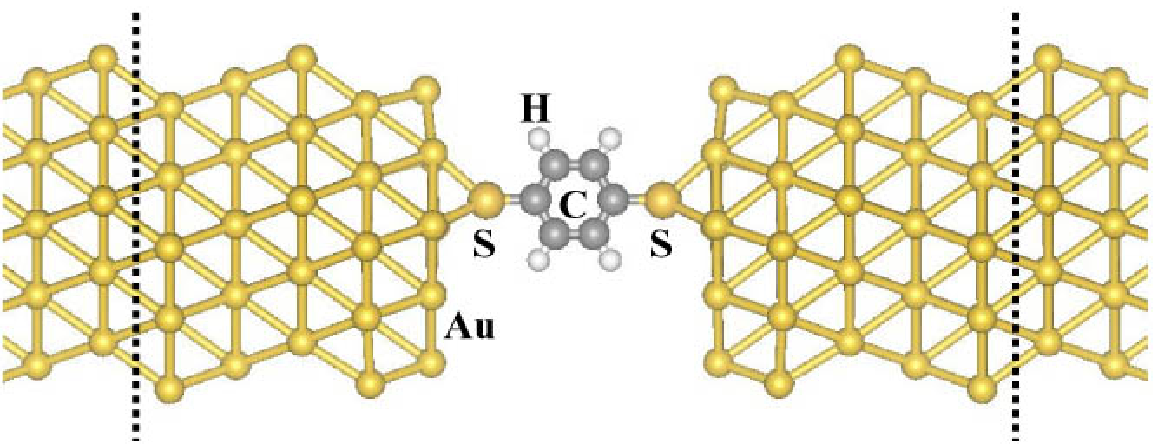} {\large \bf (d)} \\ 
\includegraphics[angle=  0,width= 7.5cm]{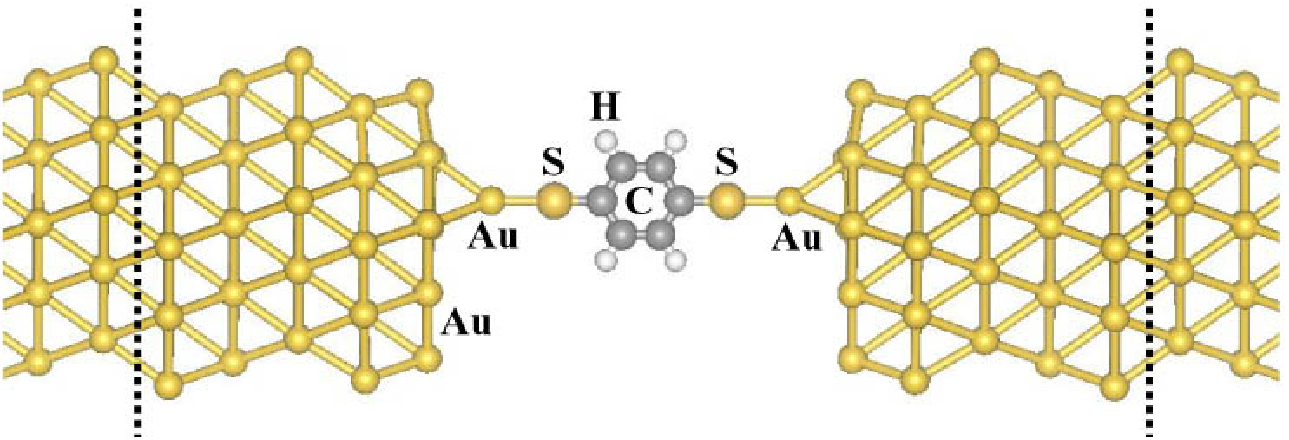} {\large \bf (e)}
\includegraphics[angle=  0,width= 7.3cm]{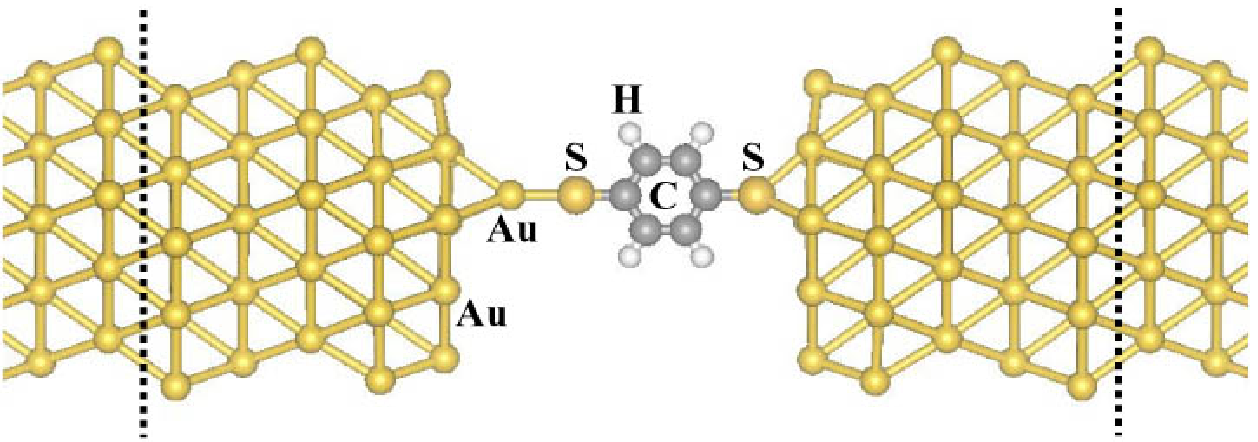} {\large \bf (f)} 
\label{fig_str}
\caption{Optimized atomic structures of the LML systems investigated, obtained by relaxing fully the molecule, the first two atomic layers of the two lead surfaces, and the molecule-lead separation.  (a), (b), and (c): The Au leads are in the (001) direction, and there are 0, 2, and 1 additional Au atoms at the contacts, respectively.  (d), (e), and (f): The Au leads are in (111) direction, and there are 0, 2, and 1 additional Au atoms at the contacts, respectively.  The dashed line indicates the interface between the device region ($C$) and the left or right lead ($L$ or $R$).  } 
\end{figure*}

%=========================== Fig. 2 =========================
\begin{figure*}[tb]
\includegraphics[angle=  0,width=15.0cm]{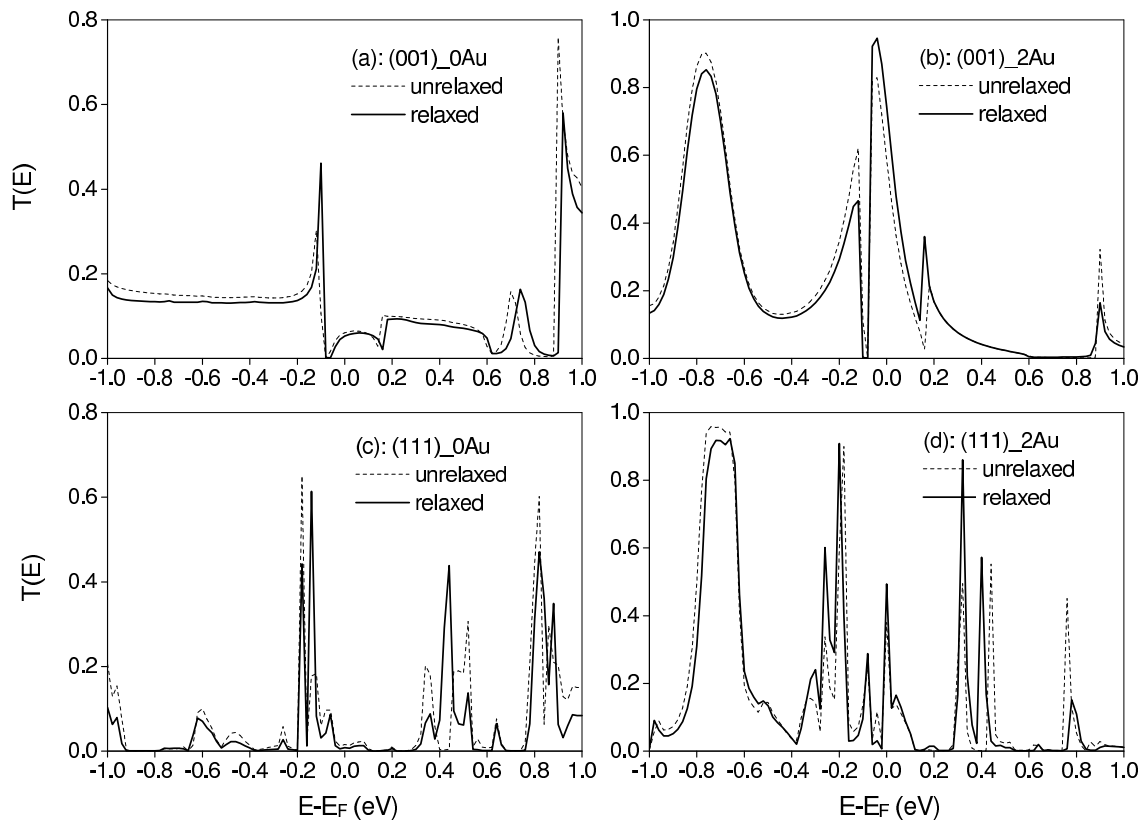}
\label{fig_t_s6}
\caption{Comparison between the transmission functions of the unrelaxed (dashed line) 
and the relaxed (solid line) systems. Different systems are indicated by 
their structure labels as defined in the text.
(a), (b), (c), and (d) correspond to the atomic structures 
of (a), (b), (d), (e) in Fig. 1, respectively. 
}
\end{figure*}

Based on this method, here we report an investigation of the molecular conductance of benzene connected to two Au leads with finite cross-section. We note that the effetcs of contact atomic structure in an unrelaxed S-anchored Au(111) system have been investigated by Xue and Ratner using a cluster approach \cite{xue2,xue3}.  We here focus on the effects of changing the atomic structure around the contacts, including the presence of an additional Au atom and changes in the molecule-lead separation, both of which are simple but important situations in break-junction experiments.
In our calculation the effects of contact atomic relaxation and different lead orientations are fully considered. Also considered are different group-VI anchoring atoms.
Our calculations show that the contact atomic structure is critically important to electron transport through the molecule: an additional Au atom at the contacts can increase the conductance by two orders of magnitude due to a LUMO-like resonance peak around the Fermi energy. We find that the dependence of the equilibrium conductance
on the molecule-lead separation can be complicated: either very weak or very strong depending on the separation regime.  The reasonability of the present systematic results are discussed in comparison with the results from Xue and Ratner \cite{xue2,xue3}.

\section{Systems investigated and computational details}

The systems we have studied consist of a benzene molecule connected to two Au leads of finite cross-section through a S atom located at the hollow site of the Au(001) or Au(111) surface. The size of the Au leads in directions perpendicular to the transport direction is set to be 4$\sqrt{2}$$\times$4$\sqrt{2}$ for the (001) lead and 3$\times$3 for the (111) lead.  To investigate the role of contact atomic structure, here we consider a very simple but possible situation: the presence of an additional Au atom at either one or both contacts, denoted by 1Au and 2Au, respectively. We use the structural label (001)\_1Au, for instance, to denote the system with the Au(001) leads and an additional Au atom at one of its contacts.  To show the effect of change in the molecule-lead separation, we change rigidly (i.e., without any further structure relaxation) the contact Au-S distance ($d_{\rm Au-S}$) in the vertical direction.  The purpose of all these considerations is to simulate possible situations in break-junction experiments in which different contact atomic structures may occur because of atomic fluctuations on the break surfaces and the molecule-lead separation can be adjusted by the MCB technique.  To show the effect of contact atomic relaxation, we calculate the electron transmission for two independent cases: (1) Atoms in the leads are fixed at their bulk positions, and the molecule is fixed at the optimized structure of the isolated molecule with the dangling bond on the S atom saturated by an Au atom. The distance between the S atom and the Au surface, however, is optimized. This structure is called ``unrelaxed''.
(2) The structure of the molecule, the fist two atomic layers of the lead surfaces, as well as the molecule-lead separation are fully relaxed. The in-plane position of the S atom is, however, fixed at the hollow site.  This structure is called `'relaxed''; the structure of the relaxed systems are shown in Fig. 1.

We use the efficient full DFT package SIESTA \cite{siesta} to do the electronic structure calculation. It adopts a LCAO-like and finite-range numerical basis set and makes use of pseudopotentials for the atomic cores. We adopt a high level double zeta plus polarization (DZP) basis set for all atomic species.  The PBE version of the generalized gradient approximation (GGA) \cite{pbe} is adopted for the electron exchange and correlation, and optimized Troullier-Martins pseudopotentials \cite{tmpp} are used for the atomic cores.  The atomic structure of the relaxed systems are optimized until the maximum residual force on all atoms is less than 0.02 eV/{\AA}.

For the transport calculation, we divide an infinite LML system
into three parts: left lead ($L$), right lead ($R$), and device region ($C$) which 
contains the molecule and large parts of the left and right leads, as shown in Fig. 1,
so that the molecule-lead interactions can be fully accommodated.
Under a bias $V_b$ the region $C$ will be driven out of equilibrium. 
We have developed a simple while strict full self-consistent approach \cite{trank1} 
to handle a steady state bias:
The bias is included through the density matrix of the region $C$ ($\mathbf{D}_C$)
in the Green function calculation instead of the potential ($\mathbf{H}_C$) in the DFT part.
Specifically, we calculate $\mathbf{D}_C$ under the boundary condition that there is a potential
difference $V_b$ between the left side of region C (together with the left lead) and the right side of C
(together with the right lead),
\begin{eqnarray}
\mathbf{D}_C\!\!\!&=&\!\!\!\frac{1}{2\pi}\!\int_{-\infty}^{+\infty}
                      \!\!\!\!\!dE\left[ \label{equ_d4}
            \mathbf{G}_{C}(E)\mathbf{\Gamma}_L(E+\frac{eV_b}{2})
            \mathbf{G}_{C}^{\dagger}(E)f(E-\mu_L) \right. \nonumber \\
          &+&\left.\!\!\mathbf{G}_{C}(E)\mathbf{\Gamma}_R(E-\frac{eV_b}{2})
            \mathbf{G}_{C}^{\dagger}(E)f(E-\mu_R)\right],
\end{eqnarray}
where $\mathbf{G}_{C}(E)$ is the retarded Green function of region $C$
(in which all the potential shifts are included \cite{trank1}),
$f$ is the Fermi function, and $\mu_L$ and $\mu_R$ the chemical 
potentials of the leads.  $\mathbf{\Gamma}_{L}(E)$ and $\mathbf{\Gamma}_{R}(E)$ reflect the coupling at energy $E$ between the $C$ region and the leads $L$ and $R$.
The self-consistent loop is $\rightarrow$ $\mathbf{H}_C$(DFT) $\rightarrow$ $\mathbf{G}_C$
$\rightarrow$ $\mathbf{D}_C$(NEGF) $\rightarrow$ $\mathbf{H}_C$(DFT) $\rightarrow$ ...
until $\mathbf{H}_C$ and $\mathbf{D}_C$ converge \cite{trank1}.
The electron transmission through $C$ is then related to Green functions by
\begin{equation} \label{equ_t}
T(E,V_b)=\textrm{Tr}\left[\mathbf{\Gamma}_L(E+\frac{eV_b}{2})\mathbf{G}_{C}(E)
         \mathbf{\Gamma}_R(E-\frac{eV_b}{2})\mathbf{G}_{C}^{\dagger}(E)\right].
\end{equation}
Note how $V_b$ again appears in $\Gamma$ here. Finally, the steady-state current is obtained by simply integrating the transmission over the energy window.

\section{Results and discussion}

\subsection{Equilibrium Transmission and Electron Transfer}

In Fig. 2 we show the transmission functions for both the unrelaxed and the relaxed systems.  The calculated values of equilibrium conductance are given in Table 1, together with the molecule-lead electron transfer (a positive value means electrons are transfered from the Au leads to the molecule, including the two S atoms).

%============================== Table 1 ========================
\begin{table}[tb]
\caption{Calculated equilibrium conductance ($G$, in units of $2e^2/h$) and 
molecule-lead electron transfer ($\Delta Q$, in units of electron, a positive value means that electrons are transferred from lead to molecule). Note the large effect of adding additional Au atoms.
}
\begin{ruledtabular}
\begin{tabular}{llldddd}
 & & & \multicolumn{2}{c}{unrelaxed} & \multicolumn{2}{c}{relaxed} \\
 & & & \multicolumn{1}{c}{$\Delta Q$} & \multicolumn{1}{c}{$G$} 
         & \multicolumn{1}{c}{$\Delta Q$} & \multicolumn{1}{c}{$G$} \\
\hline
S & (001) &  0Au & -0.026 & 0.061 & -0.048 & 0.053 \\
  &       &  1Au &        &       & +0.169 & 0.059 \\
  &       &  2Au & +0.200 & 0.590 & +0.261 & 0.740 \\

  & (111) &  0Au & +0.044 & 0.016 & +0.053 & 0.0080 \\
  &       &  1Au &        &       & +0.204 & 0.025 \\
  &       &  2Au & +0.178 & 0.380 & +0.228 & 0.490 \\

\end{tabular}
\end{ruledtabular}
\end{table}

Because we use the bulk Au structure for the leads in the unrelaxed cases, the contact atomic relaxation consists of two parts: (1) the relaxation of the bare Au lead with respect to its bulk structure, and (2) the relaxation of both the leads and the molecule induced by the molecule-lead interaction.  Our calculations of optimized atomic structure show that both parts are very small, as can be seen in Fig. 1. The small relaxation of the bare Au leads is consistent with the very small surface relaxation of unreconstructed infinite Au surfaces. The small molecule-lead relaxation is understandable because the hollow site corresponds to a bulk atomic position, and so the absorption of a S atom will not markedly change the directional binding of the surface. Because the contact atomic relaxation is very small, its effect on electron transmission is only minor: as seen in Fig. 2 and Table 1, the induced change in equilibrium conductance is less than 100\%.  The very small molecule-lead relaxation justifies the reasonability of changing rigidly $d_{\rm Au-S}$ for simulating the change in the molecule-lead separation.

The results in Table 1 show that if there is no additional Au atom at the contacts (0Au systems), the Au(001) lead always yields a larger conductance (by about 6 times) than Au(111).  This can be understood by considering the contact atomic configuration in the two cases: the S atom touches four Au atoms on the Au(001) surface but only three on Au(111). From Table 1, the two 0Au systems have very similar electron transfers (very small). In contrast, from Fig. 2 (a) and (c), their transmission functions around the Fermi energy are quite different: there is a small peak for Au(001) which is absent for Au(111). Taken together, this indicates that the large difference in conductance between the two lead orientations is mainly due to the different molecule-lead couplings.  Note that this large difference in conductance is significantly reduced by adding additional Au atoms to the contacts. This is obvious according to the above analysis: the additional Au atom reduces the structural difference for electron transport between the Au(001) and Au(111) contacts.
%************************************************************************************
Another difference between the two lead orientations is that 
the overall structure of T(E) is totally different, as shown in Fig. 2: 
The T(E) functions of the Au(001) systems are quite smooth while those of the 
Au(111) systems have many very sharp structures. We cannot understand well at this moment 
this difference. Possible reasons include
the too thin Au(111) lead and its lower symmetry. This behavior of the Au(111) lead has
also been found in other calculations \cite{transiesta}.
%************************************************************************************

%=========================== Fig. 3 =========================
\begin{figure}[tb]
\includegraphics[angle=  0,width= 5.5cm]{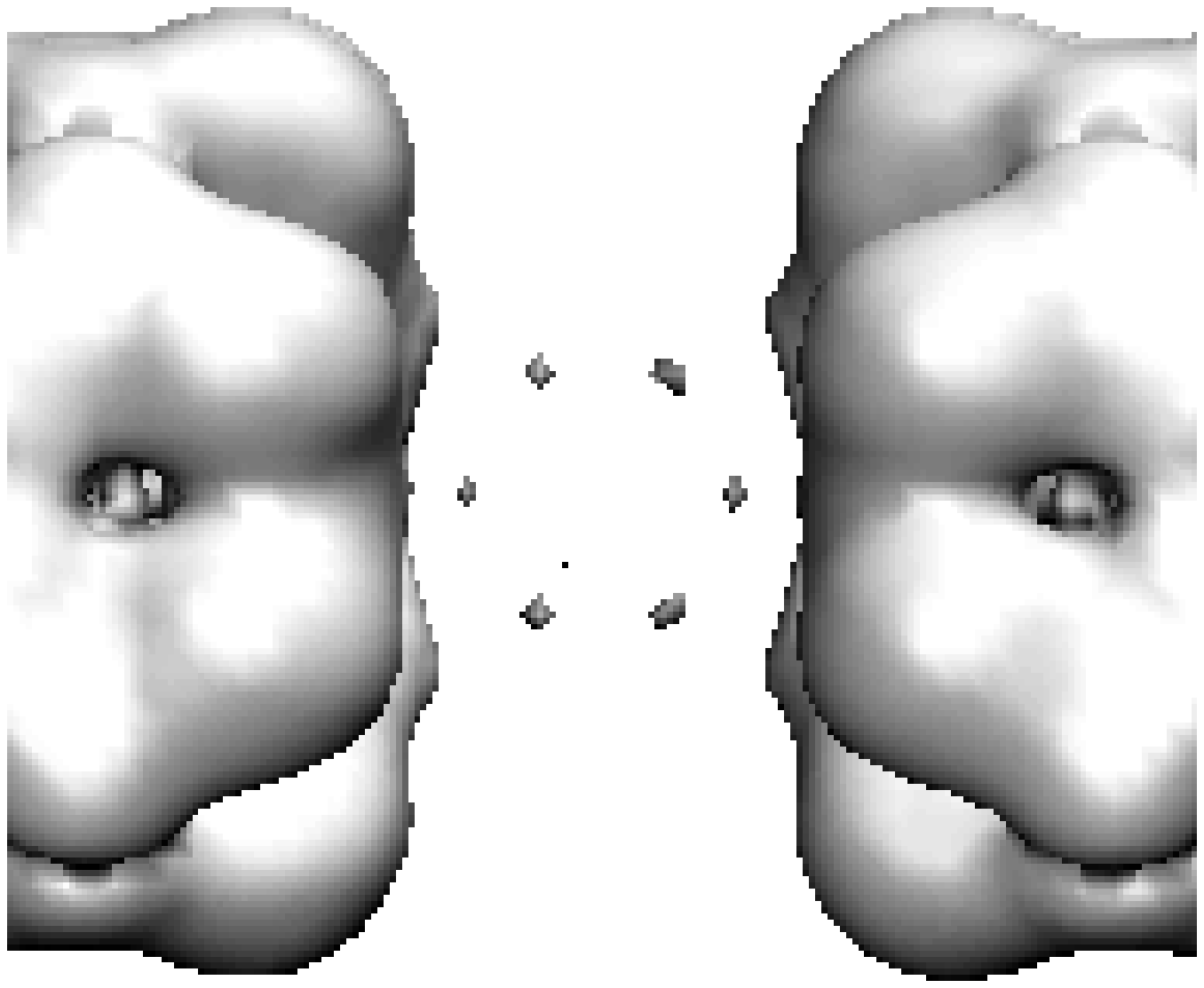}
\includegraphics[angle=  0,width= 7.0cm]{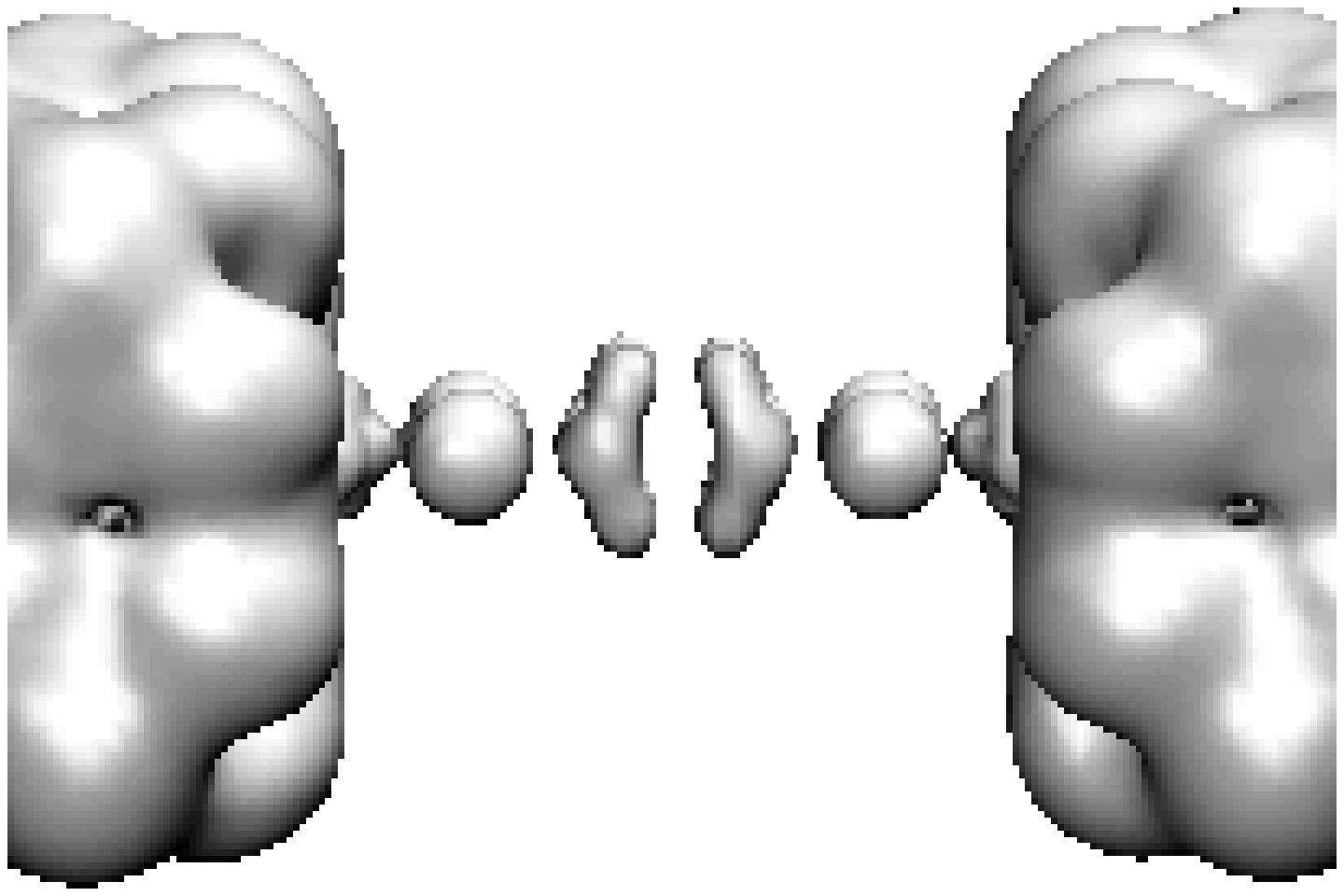}
\label{ldos}
\caption{Local density of states within the energy window [-0.05, +0.05] eV
around the Fermi energy for the systems (001)\_0Au (upper) and 
(001)\_2Au (lower). Note the channel with LUMO-like character in the latter which is absent in the former.
}
\end{figure}

The introduction of Au atoms to the contacts causes a dramatic change in the conductance of the system. The calculated values of equilibrium conductance in Table 1 show that an additional Au atom at both contacts increases the conductance by a factor of 14 for the Au(001) lead and 61 for Au(111).  From the calculated transmission functions in Fig. 2, we see clearly that the two additional Au atoms produce a large resonance around the Fermi energy. It will be shown later that this originates from the LUMO level of the isolated molecule. The driving forces causing the resonance peak are the molecule-lead 
%\marginpar{\bf ??}
electron transfer and coupling. As seen in Table 1, the introduction of the two additional Au atoms changes the sign of the electron transfer so that more electrons are transfered from the leads to the molecule, finally causing the (broadened) molecular LUMO level to line up with the chemical potential of the leads. This point will be confirmed directly later.  To show more clearly the difference between the systems with and without the additional
Au atoms we show in Fig. 3 the local density of states within the energy window [-0.05, +0.05] around the Fermi energy for the systems (001)\_0Au and (001)\_2Au. The two additional Au atoms clearly lead to a conductance channel in the (001)\_2Au system which is absent in the (001)\_0Au case. Note that the spatial shape of the density states on the molecule indicates that the channel is formed from the LUMO level. When we add the Au to only one of the two contacts [see the atomic structures in Fig. 1 (c) and (f)], the increase in conductance is only minor (see the rows with label ``1Au'' in Table 1), especially for the Au(001)-lead, indicating that the electron transmission through the molecule is then choked off by the less transparent contact.

%============================== Table 2 ========================
\begin{table}[tb]
\caption{Calculated equilibrium conductance ($G$, in units of $2e^2/h$) and molecule-lead electron transfer (($\Delta Q$, in units of electron) for the Se- and Te-
anchored systems. (The notations are the same as those in Table 1.) All the trends seen in the S anchored systems are also evident here.
}
\begin{ruledtabular}
\begin{tabular}{llldddd}
 & & & \multicolumn{2}{c}{unrelaxed} & \multicolumn{2}{c}{relaxed} \\
 & & & \multicolumn{1}{c}{$\Delta Q$} & \multicolumn{1}{c}{$G$} 
         & \multicolumn{1}{c}{$\Delta Q$} & \multicolumn{1}{c}{$G$} \\
\hline
Se& (001) &  0Au & -0.190 & 0.036 & -0.206 & 0.031 \\
  &       &  1Au &        &       & +0.099 & 0.044 \\
  &       &  2Au & +0.220 & 0.490 & +0.283 & 0.660 \\

  & (111) &  0Au & -0.042 & 0.010 & -0.010 & 0.0047 \\
  &       &  1Au &        &       & +0.167 & 0.036 \\
  &       &  2Au & +0.198 & 0.357 & +0.235 & 0.550 \\

\hline
Te& (001) &  0Au & -0.318 & 0.017 & -0.294 & 0.014 \\
  &       &  1Au &        &       & +0.052 & 0.024 \\
  &       &  2Au & +0.228 & 0.260 & +0.290 & 0.420 \\

  & (111) &  0Au & -0.088 & 0.0050& -0.044 & 0.0024 \\
  &       &  1Au &        &       & +0.155 & 0.022 \\
  &       &  2Au & +0.216 & 0.240 & +0.261 & 0.430 \\

\end{tabular}
\end{ruledtabular}
\end{table}

%=========================== Fig. 4 =========================
\begin{figure}[tb]
\includegraphics[angle=  0,width= 8.5cm]{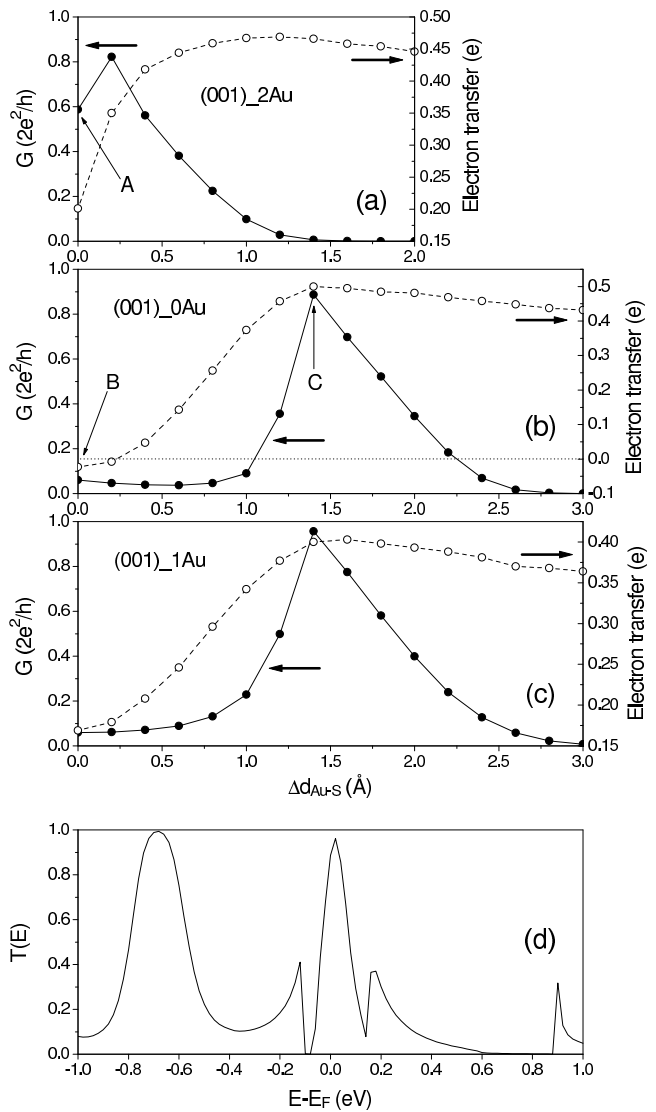} 
\label{fig_t-d3}
\caption{Equilibrium conductance and lead-to-molecule electron transfer as functions of the change in Au-S distance ($\Delta d_{\rm Au-S}$), for (a) unrelaxed (001)\_2Au system, (b) unrelaxed (001)\_0Au system, and (c) relaxed (001)\_1Au system.  In (a) and (b) $d_{\rm Au-S}$ of the two contacts is rigidly and symmetrically  changed, while in (c) only $d_{\rm Au-S}$ of the contact without the additional Au is rigidly changed.  The transmission function for point C is shown in (d), while those for points A and B are already shown in Fig. 2(b) and (a), respectively.  Note the large resonance conductance peak in panels (a)-(c) accompanied by significant electron transfer to the molecule. The similarity of (d) to Fig. 2(b) shows that increased Au-S separation has an effect comparable to that of the extra Au atoms.
} 
\label{Gvsd}
\end{figure}

\subsection{Other Anchoring Atoms}

In order to examine whether this behavior is common for other group-VI anchoring atoms, we carry out the same calculation for systems having similar geometry but anchored by Se and Te atoms. The results of equilibrium conductance and molecule-lead electron transfer are listed in Table 2. Clearly, the conclusions reached in the S-anchored systems also holds for the Se- and Te-anchored systems.

\subsection{Dependence on Molecule-Lead Separation}

In break-junction experiments the molecule-lead separation may not be at its equilibrium value but rather is probably lengthened or compressed because of the mismatch between the molecular length and the junction break. To simulate this situation, here we calculate the equilibrium conductance as a function of the change in $d_{\rm Au-S}$ ($\Delta d_{\rm Au-S}$) with regard to its equilibrium value for three systems: unrelaxed (001)\_2Au, (001)\_0Au, and relaxed (001)\_1Au. For the first two cases, $d_{\rm Au-S}$ of both contacts will be changed rigidly while maintaining the symmetry of the system, while for the third case only $d_{\rm Au-S}$ of the contact without the additional 
%\marginpar{\bf why??}
Au atom will be changed rigidly.  As mentioned previously, the very small molecule-lead relaxation justifies this treatment.  The results are shown in Fig.~\ref{Gvsd}. 

There is a large resonance peak in the conductance curve for all three systems. For the (001)\_2Au case, the equilibrium Au-S distance is very close to the position of the resonance peak, while for the other two systems the equilibrium Au-S distance is about 1.4{\AA} away from the position of the resonance peak.  Along with the increase of $\Delta d_{\rm Au-S}$ the amount of electron transferred from the leads to the molecule increases and reach its maximum around the resonance peak.  If we assume the mechanism producing the resonance is the same for the three systems, then $T(E)$ at point C in Fig.~\ref{Gvsd}(b) should be similar to that at point A in Fig.~\ref{Gvsd}(a), even though transmission at points A and B are very different [see Fig. 2(b) and (a)].  To check our point we plot the transmission function for point C in Fig.~\ref{Gvsd}(d). It is clear that, as expected, this $T(E)$ is quite similar to that of Fig. 2(b).

Before continuing to investigate the mechanism of the resonance, we briefly pause to compare to a previous calculation: the S-anchored Au(111)-lead system was investigated using a cluster method \cite{xue2, xue3} which did not, however, include the atomic relaxation of the contact. In the area where they overlap, there is qualitative agreement between the two calculations; that is, they both find that the presence of the additional Au atoms at the contacts significantly increases the equilibrium conductance and that this effect is quite similar to that from increasing the contact Au-S distance. However, there are significant quantitative differences between the two sets of results.  For instance, the gap in $T(E)$ for the (111)\_0Au case in Ref.\ \  \onlinecite{xue2} is much larger than our result [compare Fig. 9 in Ref.\ \  \onlinecite{xue2} to Fig. 2(c) here]; the latter is comparable to the GGA HOMO-LUMO gap of the molecule ($\sim$ 0.3eV) and thus appear more reasonable physically.  These significant differences are probably caused by the very different techniques used in the two calculations: (1) In our method the electronic structure of the molecule and the leads are treated {\it exactly} on the same footing \cite{test1} while in Ref.\ \ \onlinecite{xue2, xue3} the molecule is treated by DFT but the leads are treated by a tight-binding approach. (2) We use periodic boundary conditions with large
parts of the leads included in the device region; as a result, the semi-infinite leads and the infinite LML system are nearly perfect in geometry \cite{test2}.  In contrast, in Ref.\ \ \onlinecite{xue2, xue3} a cluster geometry is used for the device region, introducing some surface effects. Our consistent treatment of the semi-infinite leads also allows us to analyze further the mechanism of the resonance peaks, as shown in the next section.

\subsection{Resonance Mechanism}

%=========================== Fig. 5 =========================
\begin{figure}[tb]
\includegraphics[angle=  0,width= 8.5cm]{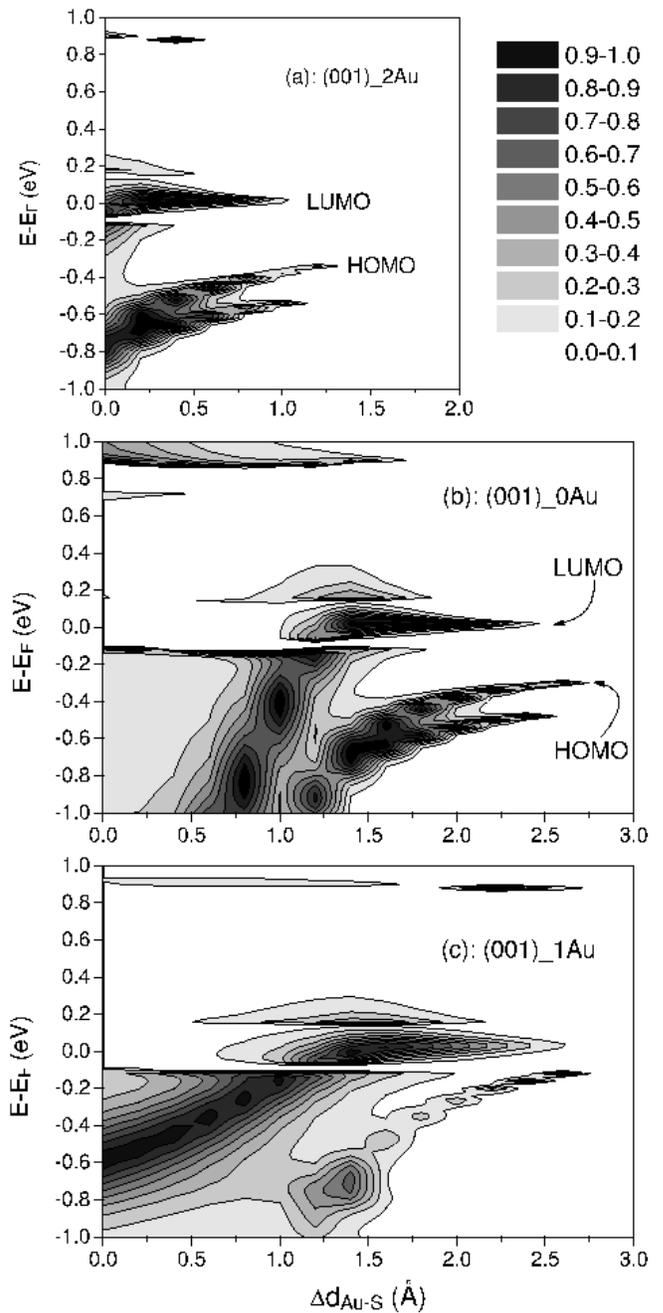}
\label{fig_t-map3}
\caption{Contour plots of transmission as a function of both energy and $\Delta d_{\rm Au-S}$ for (a) unrelaxed (001)\_2Au system, (b) unrelaxed (001)\_0Au system, and (c) relaxed (001)\_1Au system. The meaning of $\Delta d_{\rm Au-S}$ is the same as in Fig.~\ref{Gvsd}. The contributions from the HOMO and LUMO orbitals of the isolated molecule are indicated in (a) and (b).  } \end{figure}

In order to show more clearly the mechanism of the resonance peak around the Fermi energy and the modification in transport properties caused by changing the contact Au-S distance, we show in Fig.~5 contour plots of transmission coefficient as functions of energy and $\Delta d_{\rm Au-S}$ for the three S-anchored systems, (001)\_2Au, (001)\_0Au, and (001)\_1Au.  For the (001)\_2Au system, when the Au-S distance is large ($\Delta d_{\rm Au-S}$ $>$ 1.0{\AA}) there are three energies in the energy window which contribute to the electron transmission.  Comparison with the level structure of the isolated molecule shows that these three energies correspond to the LUMO, HOMO, and HOMO-1 of the isolated molecule, respectively. As $\Delta d_{\rm Au-S}$ decreases, the increasing molecule-lead coupling broadens these three levels. The two HOMO states shift gradually to lower energies as well.  It is clear that the resonance peak around $\Delta d_{\rm Au-S}$ = 0.2{\AA} in Fig.~\ref{Gvsd}(a) originates from the (broadened) LUMO contribution.

For the (001)\_0Au case and large Au-S distance ($\Delta d_{\rm Au-S}$ $>$ 1.4{\AA}), the result is completely similar to that for the (001)\_2Au system. This is a further indication that the two resonance peaks in Fig.~\ref{Gvsd} (a) and (b) have the same character,  as we already argued from the similarity similarity $T(E)$ in Fig. 2(b) and Fig.~\ref{Gvsd}(d). The role of the two additional Au atoms at the contacts is equivalent to that from increasing the surface-S distance.  As $\Delta d_{\rm Au-S}$ is decreased to smaller than 1.0{\AA} the strong molecule-lead coupling changes significantly the local electronic structure, and we cannot distinguish the individual contribution from the molecular orbitals anymore.  Finally, at $\Delta d_{\rm Au-S}$ = 0.0{\AA} the transmission function becomes totally different from that for large $\Delta d_{\rm Au-S}$, as shown in Fig. 2(a).  

An interesting thing we should notice in Fig.~5(b) is that within the large range of $\Delta d_{\rm Au-S}$ $\sim$ 0-1{\AA} the equilibrium conductance is actually very insensitive to the Au-S distance. This indicates that the molecule-lead separation dependence of the equilibrium conductance can be quite complicated: it can be either very strong (for large $\Delta d_{\rm Au-S}$) or very weak (for small $\Delta d_{\rm Au-S}$).  

For the (001)\_1Au system, because of the strong coupling on the left side [see Fig. 1 (c)] we cannot recognize the individual contributions from the molecular orbitals in Fig.~5(b) even for large $\Delta d_{\rm Au-S}$. However, there is a similar LUMO-like contribution to the resonance peak in Fig.~\ref{Gvsd}(c), and the equilibrium conductance is also insensitive to the Au-S distance for $\Delta d_{\rm Au-S}$ $\sim$ 0-1{\AA}. This indicates that the total electron transmission is dominated by the weakly coupled contact.

\subsection{I-V Curve}

%=========================== Fig. 6 =========================
\begin{figure}[tb]
\includegraphics[angle=  0,width= 8.5cm]{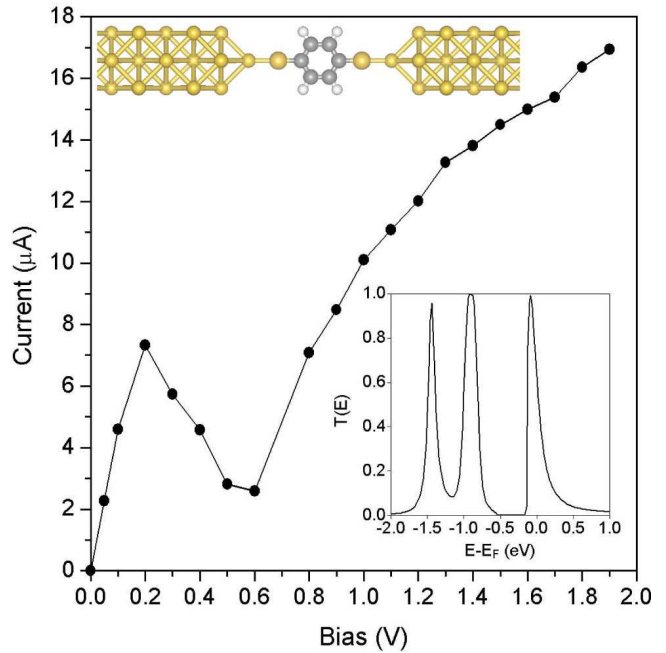}
\label{fig_iv}
\caption{$I$-$V$ curve of the smaller (001)\_2Au system shown in the upper inset, whose transmission function under zero bias is shown in the lower inset.  Note that there is a large resonance peak in $T(E)$ around the Fermi energy which causes a large negative differential conductance in the $I$-$V$ curve around $V_b$ = 0.2-0.5 V. } 
\end{figure}

The large resonance peak in the equilibrium transmission function around the Fermi energy implies a large negative differential conductance when a bias voltage is applied. We would like to show this explicitly.  In order to avoid the large computational effort for the $I-V$ curve of this system, we with a similar but slightly smaller system: two smaller 2$\sqrt{2}\times2\sqrt{2}$ Au(001) leads instead.  The structure and equilibrium transmission function of the small system are shown in Fig.~6. It can been seen that its transmission function is somewhat different from that of the large system, but the feature of the large resonance peak around the Fermi energy is the same.  The calculated $I-V$ curve given in Fig. 6 shows clearly a large negative differential conductance around $V_b$ $\sim$ 0.2 -- 0.5 V.

\section{Summary}

By using a density functional theory calculation for molecular electronic structure and a Green function method for electron transport, we have  calculated from first principles the molecular conductance of benzene sandwiched between two Au leads in different ways. In our calculation, the effects of contact atomic relaxation, two different lead orientations, and different anchoring atoms are fully considered. We focused on the effects of the change in atomic structure around the contacts, including the presence of an additional Au atom and changes in the molecule-lead separation, as an effort to simulate possible situations in break-junction experiments. Our calculations clearly show the critical role of the contact atomic  structure in electronic transport through molecules:

(1) The presence of an additional Au atom at each of the two contacts can increase the  equilibrium conductance by one to two orders of magnitude {\it regardless of} the contact atomic structure or group-VI anchoring atom. 
The mechanism is the creation of a LUMO-like resonance peak around the Fermi energy, which also leads to negative differential conductance under applied bias.

(2) The presence of the additional Au atom at only one contact will give only a minor increase in conductance because the electron transmission 
is then choked off by the other non-transparent contact. 

(3) Because of the different molecule-lead coupling, the Au(001) lead will lead to an equilibrium conductance up to 6 times larger than that of the Au(111) lead. This difference is significantly reduced by adding the additional Au atom at the contacts.

(4) The dependence of the equilibrium conductance on the molecule-lead separation is subtle: it can range from either very weak to very strong depending on the separation regimes. 

\acknowledgments
This work was supported in part by the NSF (DMR-0103003).


\begin{references}
\bibitem{test1}  In order to reduce the computational cost, we have tried to use DZP basis
                 set for the molecule but Single Zeta (SZ) basis set for the leads. We found
                 remarkable difference from the result of the all-DZP calculation.
\end{references}
\end{document}